\newif\ifdraft   \draftfalse
\newif\iffull    \fulltrue
\newif\iflater   \laterfalse  
\newif\iffoot    \footfalse   
\newif\ifnever   \neverfalse  

\makeatletter \@input{texdirectives} \makeatother

\iffull \foottrue \fi

\iffull
\documentclass[11pt]{article}
\usepackage{fullpage}
\else
\documentclass[conference]{ieeetran}
\usepackage{balance}
\fi

\usepackage{xspace}
\usepackage[usenames]{xcolor}
\usepackage{mathpartir}
\usepackage{multirow}
\usepackage{mathptmx}
\usepackage{url}
\usepackage{amsthm}
\usepackage{amsmath}
\usepackage{amssymb}
\usepackage{alltt}
\usepackage{fancyhdr}
\usepackage{setspace}
\usepackage{subfig}
\usepackage{tikz}
\usepackage{balance}

\usepackage{mathrsfs}
\usepackage{epsfig}
\usepackage{hyperref}

\newtheorem{theorem}{\noindent {Theorem}}{}
{}
\newtheorem{definition}[theorem]{\noindent {Definition}}{}
\newtheorem{proposition}[theorem]{\noindent {Proposition}}{}
\newtheorem{remark}[theorem]{\noindent {Remark}}{}
\newcommand{\mg}[1]{\ifdraft \textcolor{blue}{[MG: #1]}\fi}
\newcommand{\jh}[1]{\ifdraft \textcolor{red}{[JH: #1]}\fi}
\newcommand{\bcp}[1]{\ifdraft \textcolor{purple}{[BCP: #1]}\fi}

\newcommand{\ar}[1]{\ifdraft \textcolor{orange}{[AR: #1]}\fi}

\newcommand{\EQREF}[1]{\eqref{#1}}

\newcommand{\sleq}{\; \leq \;}
\newcommand{\sgeq}{\; \geq \; }

\newcommand{\SUBSECTION}{\subsection}
\newenvironment{mathdisplayfull}{\iffull \[ \else $ \fi}{\iffull \]\else $\ignorespaces\fi}
\newcommand{\shortbreak}{{\iffull\else\\\fi}}

\hypersetup{
      bookmarks=true,
      colorlinks=true,
      linkcolor=blue,
      citecolor=red,
      filecolor=black,
      urlcolor=blue,
      pdfauthor = {Many authors},
      pdftitle = {Working Title},
      pdfkeywords = {Keywords}
}

\begin{document}
\title{\LARGE Differential Privacy: An Economic Method for Choosing Epsilon
  \vspace{-0.7ex}}
\iffull
\author{
Justin Hsu
\quad
Marco Gaboardi
\quad
Andreas Haeberlen
\quad
Sanjeev Khanna
\\[1ex]
Arjun Narayan
\quad
Benjamin C. Pierce
\quad
Aaron Roth
  }
\fi

\maketitle

\sloppy

\begin{abstract}

\emph{Differential privacy} is becoming a gold standard for privacy research; it
offers a guaranteed bound on loss of privacy due to release of query results,
even under worst-case assumptions.  The theory of differential privacy is an
active research area, and there are now differentially private algorithms for a
wide range of interesting problems. 

However, the question of when differential privacy works \emph{in practice} has
received relatively little attention.  In particular, there is still no rigorous
method for choosing the key parameter $\epsilon$, which controls the crucial
tradeoff between the strength of the privacy guarantee and the accuracy of the
published results.

In this paper, we examine the role that these parameters play in concrete
applications, identifying the key questions that must
be addressed when choosing specific values.  This choice requires balancing the
interests of two different parties: the data {\em analyst} and the prospective
{\em participant}, who must decide whether to allow their data to be included in
the analysis. We propose a simple model that expresses this balance as formulas
over a handful of parameters, and we use our model to choose $\epsilon$ on a series of simple
statistical studies.  We also explore a
surprising insight: in some circumstances, a differentially private study can be
\emph{more accurate} than a non-private study for the same cost, under our
model. Finally, we discuss the simplifying assumptions in our model and
outline a research agenda for possible refinements.

\end{abstract}

\ifdraft
\jh{TODOs}

\begin{itemize}
  \item General polishing
  \item Double check all numbers
\end{itemize}
\fi




\section{Introduction} \label{sec:introduction}

Protecting privacy is hard: experience has repeatedly shown that when owners of
sensitive datasets release derived data, they often reveal more information than
intended. Even careful efforts to protect privacy often prove inadequate---a
notable example is the Netflix prize competition, which released movie ratings
from subscribers.  Although the data was carefully anonymized, Narayanan and
Shmatikov were later able to ``de-anonymize'' some of the private
records~\cite{narayanan-2008-deanonymization}.

Privacy breaches often occur when the owner of the dataset uses an incorrect
threat model---e.g., they make wrong assumptions about the knowledge available
to attackers. In the case of Netflix, Narayanan and Shmatikov  had access to
auxiliary data in the form of a public, unanonymized data set (from IMDB) that
contained similar ratings.  Such errors are difficult to prevent, since this requires
reasoning about {\em all} the information that could be (or become) available to an
attacker.

One way through this dilemma is to make sure that every computation on sensitive
data satisfies \emph{differential privacy}~\cite{dwork-2006-sensitivity}. This
gives a very strong guarantee: if an individual's data is used in a
differentially private computation, the probability of any given result changes
by at most a factor of $e^\epsilon$, where $\epsilon$ is a parameter controlling
the tradeoff between privacy and accuracy.  Differential privacy impresses by
the long list of assumptions it does {\em not} require: it is not necessary to
know what information an attacker has, whether attackers are colluding, or what
the attackers are looking for.

But there is one question that users of differential privacy cannot avoid:
how to choose the privacy parameter $\epsilon$.  It is the central parameter
controlling strength of the privacy guarantee, and hence the number of queries that can be
answered privately as well as the achievable accuracy. But $\epsilon$ is also a
rather abstract quantity, and it is not clear how to choose an appropriate value
in a given situation. This is evident in the
literature\iffull~\cite[etc.]{MM10,CM08,MKAGV08,KKMN09,BLST10,
  MKS11,BXCF12,LQSC12,NP12,CSS12,AC12,USF12,XXFG12,LM12,CFD11,CPSSY11,CMS09}\fi{},
where algorithms have been evaluated with $\epsilon$ ranging from as little as
0.01 to as much as 7, often with no explanation or justification.
A similar concern applies to a second parameter $\delta$ in $(\epsilon,
\delta)$-differential privacy, a standard generalization of differential
privacy~\cite{dwork2006our}.

In this paper, we take a step towards a more principled approach by examining
the impact of $\epsilon$
and $\delta$ on the different actors in a differentially private
study: the data analyst, and the prospective participants who contribute private
data.  We propose a simple model that can be used to calculate a range of
acceptable values of $\epsilon$ and $\delta$, based on a few parameters of the
study.  Our model assumes that the participants are rational and will choose to
contribute their data if their expected benefits (i.e., monetary
compensation)
from the study outweigh the risks (i.e., the bad events that may befall them as
a result of their private data being exposed).

To demonstrate our model, we use it to choose $\epsilon$ in a series of case
studies.  We start by presenting the different
parameters of our model, in the simplest situation where the analyst is
interested in the result of only one query. Then, we consider a more realistic
setting where the analyst wants to answer thousands of queries. Next, we show how
our model can incorporate constraints specific to a particular study.
Finally, we apply our model to a more accurate study under
$(\epsilon,\delta)$-differential privacy.
Throughout these examples, we vary the input parameters to our model
through four scenarios---a clinical study of smokers, a study of educational
data, a study of movie ratings, and a social network study---and show how the
conclusions of our model change.

We also find that---somewhat counterintuitively---a study with strong
differential privacy guarantees can sometimes be \emph{cheaper} or (given a
fixed budget) \emph{more accurate} than an equivalent study without any privacy
protections: while a differentially private study requires considerably more
participants to account for the additional noise, it substantially reduces the
risks of each participant and thus lowers the compensation that rational
participants should demand.

Our model provides a principled way to choose reasonable values for
$\epsilon$ and $\delta$
based on parameters with more immediate connections to the real world.
For many applications of differential privacy, this level of guidance may
already be useful.
However, like any model, ours relies on some simplifying
assumptions; for instance, we assume that participants fear some specific bad
events when participating in the study, and that they can estimate their
expected cost from these events even when they do not participate in the study.
Some applications may require a more detailed model, and we consider possible
refinements.

Our main contributions are: (1) a principled approach to choosing the privacy
parameter $\epsilon$ for differentially private data analysis
(Section~\ref{sec:model}); and (2) three
\mg{Not sure if writing four or just rephrase the next few sentences.}%
case studies: a simple one-query
study, a more sophisticated study answering many queries
(Section~\ref{sec:ex-simple}), and a study with external constraints
(Section~\ref{sec:irqdb}); and (3) an extension of our model to
$(\epsilon,\delta)$-differential privacy (Section~\ref{sec:delta}).
As an application of our model, we consider when a differentially private study
can be cheaper than a non-private study (Section~\ref{sec:cost-of-privacy}).
We discuss possible extensions of our model in Section~\ref{sec:hard}, and review
related work in Section~\ref{sec:relatedwork}.
\iffull
Finally, we conclude in Section~\ref{sec:conclusion}.
\fi

\section{Background: Differential privacy}\label{sec:background}

Before describing our model, let us briefly review the core definitions of
$\epsilon$-differential privacy.
%
(We defer the generalization of $(\epsilon,\delta)$-differential privacy to
Section~\ref{sec:delta}.)

{Differential privacy}~\cite{dwork-2006-sensitivity} is a quantitative notion of
privacy that bounds how much a single individual's private data can contribute to a
public output.
The standard setting involves a {\em database} of private information and a
{\em mechanism} that calculates an output given the database. More formally,
a {database} $D$ is a multiset of records belonging to some \emph{data
  universe} $\mathcal{X}$, where a record corresponds to one individual's
private data.  We say that two databases are \emph{neighbors} if they are
the same size and identical except for a single record.\footnote{%
  The standard definition of differential privacy~\cite{dwork-2006-sensitivity} is slightly
  different: it says that neighboring databases are identical, except one has an
  {\em additional} record. We use our modified definition since we will assume
  the database size is public, in which case neighboring databases have the {\em
    same} size.}
A {mechanism} $M$ is a randomized function that takes the database as input and
outputs an element of the range $\mathcal{R}$.

\begin{definition}[\cite{dwork-2006-sensitivity}] \label{def:diff-priv}
  Given $\epsilon \geq 0$, a mechanism $M$ is \emph{$\epsilon$-differentially
  private} if, for any two neighboring databases $D$ and $D'$ and for any subset $S
  \subseteq \mathcal{R}$ of outputs,
  \begin{equation} \label{eq:dp-ubound}
    \Pr [ M(D) \in S ] \sleq e^{\epsilon} \cdot \Pr [ M(D') \in S ].
  \end{equation}
\end{definition}
Note that $S$ in this definition is any subset of the mechanism's range. In
particular, when $S$ is a singleton set $\{s\}$, the definition states that the
probability of outputting $s$ on a database $D$ is at most $e^\epsilon$ times
the probability of outputting $s$ on any neighboring database $D'$.

For an intuitive reading of Definition~\ref{def:diff-priv}, let $x$ be an
individual in database $D$, and let $D'$ contain the same data as $D$ except
with $x$'s data replaced by default data.
Then, the differential privacy guarantee states that the probability of any
output of mechanism $M$ is within an $e^\epsilon$ multiplicative factor whether
or not $x$'s private data is included in the input.  Hence, the parameter
$\epsilon$ controls how much the distribution of outputs can depend on data from
the individual $x$.

The definition also implies a lower bound: swapping $D$ and $D'$ yields
$e^\epsilon \cdot \Pr [ M(D) \in S ] \geq \Pr [ M(D') \in S ]$, or
\begin{equation} \label{eq:dp-lbound}
  \Pr [ M(D) \in S ] \sgeq e^{-\epsilon} \cdot \Pr [ M(D') \in S ].
\end{equation}
That is, the probability of an output in $S$ on a database $D$ is \emph{at
least} $e^{-\epsilon}$ times the probability of an output in $S$ on a
neighboring database $D'$.\footnote{For example, if $\Pr [ M(D) \in S ]
  = 0$ for some $D$ and $S$, then $\Pr [ M(D') \in S ] = 0$ for \emph{all}
  databases $D'$---if some outputs are impossible on one input database, they
  must be impossible on {all} inputs. }{}


\iffull
\SUBSECTION{The Laplace and exponential mechanisms}
\fi
The canonical example of a differentially private mechanism is the \emph{Laplace
  mechanism}.

\begin{theorem}[\cite{dwork-2006-sensitivity}]
  Suppose $\epsilon, c > 0$. A function $g$ that maps databases to real numbers
  is \emph{$c$-sensitive} if $|g(D) - g(D')| \leq c$ for all neighboring
  $D, D'$. For such a function, the \emph{Laplace mechanism} is
  defined by
  \begin{mathdisplayfull}
    L_{c, \epsilon}(D) = g(D) + \nu,
  \end{mathdisplayfull}
  where $\nu$ is drawn from the Laplace distribution $Lap(c/\epsilon)$, that is,
  with probability density function
  \[
    F(\nu) = \frac{ \epsilon}{2c} \exp \left( \frac{ - \epsilon |\nu| } {c}
    \right).
  \]
  This mechanism is $\epsilon$-differentially private.
\end{theorem}

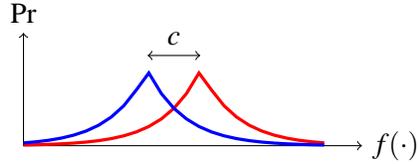
\begin{figure}
  \centering
  \begin{tikzpicture}
    \draw[<->] (-2, 1.5) node[above] {$\Pr$} -- (-2,0) -- (2.5,0) node[right]
    {$f(\cdot)$};
    \draw[very thick, color=red] plot[domain=-2:2] (\x, {1 * exp(- 2 * abs(\x -
    0.35))});
    \draw[very thick, color=blue] plot[domain=-2:2] (\x, {1 * exp(- 2 * abs(\x +
    0.35))});
    \draw[<->] (-0.35, 1.2) -- node[above] {$c$} (0.35, 1.2);
  \end{tikzpicture}
  \caption{Probability distributions of the Laplace mechanism for a
  $c$-sensitive function on two neighboring databases.}
  \label{fig:laplace}
\end{figure}

The scale $c/\epsilon$ of the Laplace distribution controls its {\em spread}:
the distribution is wider for more sensitive functions (larger $c$) or stronger
privacy guarantees (smaller $\epsilon$), giving a higher probability of adding
more noise.

For example, suppose that we have a database $D$ of medical information and we
wish to compute the proportion of smokers in a differentially private way. If
the database has $N$ records, define $g(D) = \#(\text{smokers in }D)/N$.
Notice that, on any two neighboring databases, this proportion changes
by at most $1/N$, since the numerator changes by at most $1$ if a single record is
altered.  Thus, $L(D) = g(D) + \nu$, where $\nu \sim Lap(1/N\epsilon)$ is an
$\epsilon$-differentially private mechanism.

\iffull
\SUBSECTION{Key benefits}

A key benefit of differential privacy lies in its worst-case assumptions. As
discussed above, it is difficult to reason about what auxiliary information
might be available to the adversary.  Differential privacy avoids this problem
by making {\em no} assumptions about the adversary's knowledge. Even knowing all
but one record of the database does not help the adversary learn the last
record: the output of a differentially private mechanism has approximately the
same distribution no matter what that record contains.  (Of course, in typical
scenarios, the adversary has far less information than this, but making this
worst-case assumption avoids losing sleep wondering exactly what the adversary
knows or can deduce.)

A second convenient feature of differential privacy is its flexible
framework---the statistical guarantees provided by differential privacy hold
regardless of the particular form of the records, the space of possible outputs,
and the way in which the mechanism operates internally. Furthermore, these
guarantees are preserved under \emph{arbitrary post-processing}: given a
differentially private mechanism $M$ and a function $f$ on the outputs, the
composition $f \circ M$ is differentially private. Hence, outputs of a
differentially private mechanism can be further transformed at no additional
risk to privacy.

A third useful property of differential privacy is \emph{compositionality}---the
privacy guarantee degrades gracefully when composing private mechanisms
together. For example, running $k$ $\epsilon$-differentially private
mechanisms (in series or in parallel) will yield a $k \epsilon$-private
mechanism~\cite{DRV10}.  This allows straightforward construction of more
complex algorithms out of simpler primitives, while preserving privacy.

Now, let us take a closer look at the central parameter in the definition:
$\epsilon$.

\fi

\section{Interpreting \texorpdfstring{$\epsilon$}{epsilon}} \label{sec:interp}
A natural interpretation of differential privacy is in terms of \emph{bad
events}. For concreteness, let the mechanism be a scientific study,
and suppose the individual has a choice to contribute data.\iffoot\footnote{
  In some situations, it is not natural to think of the individual as having a
  choice in participating. For instance, when an individual visits a
  website, the server can record information from this event---the individual
  may not even notice that her information is being aggregated.  However, the
  standard setting for differential privacy assumes that the individual
  can freely decide whether or not to participate, so this is a good place to
  begin; section~\ref{sec:irqdb} discusses an alternative situation.
}\fi{}
Let $\mathcal{U}$ be the space of all real-world events, and suppose
$\mathcal{E} \subseteq \mathcal{U}$ is a set of events such that, if the
output of the mechanism is fixed, an individual's participation
has no effect on the probabilities of events in $\mathcal{E}$ (we
will make this more precise below).\footnote{Events that do not
  satisfy this technical condition may not be protected by differential
  privacy; further details about this issue can be found in
  Appendix~\ref{app:protected}. }{}
Note that probabilities of events in $\mathcal{E}$ may still depend on the
output of the mechanism.  Roughly, $\mathcal{E}$ can be thought of as the set
of privacy violation events.

To connect the outputs of the mechanism to the real-world events in $\mathcal{E}$, we
imagine two runs of the mechanism: one with $x$'s real data and one with
dummy data,\iffoot\footnote{This can be thought of as a default record, or even
  random data. }\fi{}
holding the other records fixed in both runs. Let $x_p$ be the event ``$x$
participates,'' $x_{np}$ be the event ``$x$ does not participate,'' and $R$ be
the output of the mechanism (a random variable). For any event $e \in
\mathcal{E}$, the probability of  event $e$ if $x$ participates in the study is
%
\[
  \Pr[ e \mid x_p ] = \sum_{r \in \mathcal{R}} \Pr[ e \mid x_p, R = r] \cdot \Pr[ R =
  r \mid x_p].
\]
%
We say that events cannot {\em observe} $x$'s participation if
all the differences between the two trials are due to differences in the output:
if the output is the same in both trials, the probability of events in
$\mathcal{E}$ should be the same.
That is, the first probability under the summation is the same assuming event
$x_p$ or $x_{np}$.

By differential privacy
(Equation~\EQREF{eq:dp-ubound}), the second probability is bounded by
$e^\epsilon$ times the probability of output $r$ if $x$ does not participate:
\iffull
\begin{align*}
  \sum_{r \in \mathcal{R}} \Pr[ e \mid x_p, R = r] \cdot \Pr[ R = r \mid x_p]
  &= \sum_{r \in \mathcal{R}} \Pr[ e \mid x_{np}, R = r] \cdot \Pr[ R = r \mid
  x_p] \\
  &\sleq e^\epsilon \cdot \sum_{r \in \mathcal{R}} \Pr[ e  \mid  x_{np}, R = r]
  \cdot \Pr[ R = r \mid x_{np}] = e^\epsilon \cdot \Pr[ e \mid x_{np} ],
\end{align*}
\fi
where the first equality is from our assumption that events in $\mathcal{E}$
cannot observe $x$'s participation.  In particular, if $e$ is a bad event in
$\mathcal{E}$, the probability of $e$ increases by at most a factor $e^\epsilon$
when the individual participates compared to when the individual does not
participate.
Hence, the interpretation of privacy in terms of bad events: under differential
privacy, events in $\mathcal{E}$ will not increase much in probability if an
individual participates or not.\iffoot\footnote{%
  By a similar calculation applying Equation~\EQREF{eq:dp-lbound}, we have $\Pr[ e
  \mid x_p ] \geq e^{-\epsilon} \Pr [e \mid x_{np}]$.  In particular, if $e$ is a
  {\em beneficial} event in $\mathcal{E}$, this bound means that $e$ will not become
  much {\em less} likely if the individual decides to participate. }\fi{}

\iffull
  We stress an important non-benefit: differential privacy does not protect a
  participant from ever coming to harm. Indeed, it is hard to see how anything
  can---what if the event was already going to happen, even if the participant
  had declined to participate in the study? Instead, the guarantee states that
  harmful events do not become much more likely if an individual participates. For
  example, a differentially private medical study cannot promise that participants
  will continue to pay the same amount for health insurance. Rather, participating
  in the study increases the risk of a rise in premiums by at most a small factor,
  compared with declining to participate.
\fi

Since differential privacy bounds the {\em multiplicative} change in
probabilities, the probability of a likely event may change significantly in
{\em absolute} terms.  Thus, the differential privacy guarantee is
stronger for events that are very unlikely to happen if the individual does not
participate.  This is arguably true of most unpleasant events concerning private data:
for instance, the probability that an individual's genome is released if they do
not participate in a genetic study is typically low.

\SUBSECTION{Introducing cost}
Of course, not all bad events are equally harmful. To model this fact, we can
assign a cost to each event.  Specifically, suppose the potential participant
has a non-negative
\emph{event cost function} $f_{\mathcal{E}}$ on the space of events
$\mathcal{E}$. Let $R$ again be the output of mechanism $M$, and define the
associated \emph{output cost function} $f$ on the space of outputs $\mathcal{R}$
by
\[
  f(r) = \mathbb{E}_{e \in \mathcal{E}} [ f_{\mathcal{E}} (e) \mid R = r].
\]
Note that
\[
  \mathbb{E}_{e \in \mathcal{E}} [ f_{\mathcal{E}} (e) \mid x_p] = \mathbb{E}_{r
    \in \mathcal{R}} [ f(r) \mid x_p],
\]
and similarly with $x_{np}$ instead of  $x_p$, so bounds on the expected value
of $f$ carry over to bounds on the expected value of $f_{\mathcal{E}}$.  Thus,
the individual need not reason about the set of outputs $\mathcal{R}$ and the
output cost function $f$ directly; they can reason just about costs of real-world events,
represented by $f_{\mathcal{E}}$.\iffoot\footnote{This is important, since
  reasoning precisely about how outputs of a mechanism influence events in the
  real world can be very difficult. }\fi{}

Using the differential privacy guarantee, we can bound the expected cost of
participating in the study:\footnote{%
  For one direction,
  \begin{align*}
    \mathop{\mathbb{E}}_{r \in \mathcal{R}} [ f(r) \mid x_p ] &= \sum_{r \in \mathcal{R}} \Pr
    [ R = r \mid x_p ] \cdot f(r) \\
    &\sleq \sum_{r \in \mathcal{R}} e^\epsilon \Pr [ R = r \mid x_{np} ] \cdot f(r)
    = e^\epsilon \cdot \mathop{\mathbb{E}}_{r \in \mathcal{R}} [ f(r) \mid
    x_{np} ].
  \end{align*}
Note that the inequality requires $f(r) \geq 0$.  The other direction is
similar, appealing to Equation~\EQREF{eq:dp-lbound}.
}{}
\begin{equation} \label{eq:dp-costs}
  e^{-\epsilon} \mathop{\mathbb{E}}_{r \in \mathcal{R}} [ f(r) \mid x_{np} ]
  \sleq  \mathop{\mathbb{E}}_{r \in \mathcal{R}} [ f(r) \mid x_p ]
  \sleq  e^{\epsilon} \mathop{\mathbb{E}}_{r \in \mathcal{R}} [ f(r) \mid x_{np} ]
\end{equation}
In other words, the expected cost of $x$ participating in a study is within an
$e^\epsilon$ factor of the expected cost of declining.

Note that $\mathcal{E}$ and the cost function $f$ have a large impact on the
expected cost: for instance, if $\mathcal{E}$ contains bad events that will not
actually be affected by the output of the mechanism, such as the event that an
asteroid impact destroys civilization, the participant's perceived increase in
expected cost may be prohibitively (and unjustifiably) large. 

In general, the question of what events a differentially private study may be
responsible for (i.e., what events should be in $\mathcal{E}$) is not a purely
technical question,
and could conceivably be handled by the legal system---just as laws
describe who is liable for bad events, perhaps laws
could also describe which events a private mechanism is liable for. Accordingly,
our model does not specify precisely which events to put in $\mathcal{E}$, as
long as they do not depend directly on the individual's participation. For our
examples, we will consider events that clearly can result from running a private
mechanism.

\SUBSECTION{The challenge of setting \texorpdfstring{$\epsilon$}{epsilon}}

So far, we have considered what $\epsilon$ means for the participant: higher
values of $\epsilon$ lead to increases in expected cost. As we will soon see,
there is another important consideration: $\epsilon$ controls how much noise is
needed to protect privacy, so it has a direct impact on accuracy.

This is the central tension---abstractly, $\epsilon$ is a knob that trades
off between privacy and utility. However, most prior work (we discuss some
exceptions in Section~\ref{sec:relatedwork}) focuses on how
the knob works rather than how it should be set.
High-level discussions about setting
$\epsilon$ tend to offer fairly generic guidance, for example reasoning that a
$10\%$ increase in the probability of a bad event that is already very
improbable is a minor concern, so $0.1$ is a sensible value for
$\epsilon$.  On the other hand, experimental evaluations of differential
privacy, where a concrete choice of $\epsilon$ is required, often just pick a
value (ranging from $0.01$ \cite{CMS09} to $7$ \cite{MKS11}) with little
justification.

In a sense, the difficulty of choosing $\epsilon$ is a hidden consequence of a
key strength of differential privacy: its extreme simplicity. That is,
$\epsilon$ is difficult to think about precisely because it rolls up into a single
parameter a fairly complex scenario involving at least two parties
with opposing interests (the analyst and the participants), as well as
considerations like compensating individuals for their risk.

Our goal in this paper is to unpack this complexity and offer a more
ramified model with more intuitive parameters. 

\section{A two-party model}\label{sec:model}

We propose a simple model for choosing $\epsilon$, involving two rational
parties: a data analyst and an individual considering whether to participate
in the analyst's study.

\SUBSECTION{The analyst's view}

The analyst's goal is to conduct a study by running a private mechanism, in
order to learn (and publish) some useful facts.  The
analyst's main concern is the \emph{accuracy} $A_M$ of the mechanism's result, with
respect to some benchmark.

One natural benchmark is the ``true answer'' for the non-differentially-private version
of the study, which we call the {\em sample statistic}.  Compared to this
standard, the error in a private study is due entirely to noise
added to preserve privacy.  This error is determined partly by $\epsilon$, but
also can depend on $N$, the number of records in the analyst's database: if a
larger number of records leads to less privacy loss to any individual, less
noise is needed to protect privacy.\iffoot\footnote{%
  For example, if the Laplace mechanism is used to release an average value, the
  sensitivity of the underlying function depends on $N$: as $N$ increases, the
  sensitivity decreases, so less noise is needed to achieve a given level of
  privacy.
}\fi{}

Another possible benchmark is the true answer on the entire population, which
we call the {\em population statistic}. This is the natural benchmark when we
want to infer properties of the population, given only a random sample of
individual data (here, the database). For this benchmark, an additional source
of error is \emph{sampling error}: the degree to which the sample is not perfectly representative of
the population. This error tends to decrease as $N$ increases: larger samples
(databases) are more representative.  This error is not due to differential
privacy and so is independent of $\epsilon$.

Since these errors typically decrease as $N$ increases, the analyst might like to conduct
huge studies, were it not for a second constraint: \emph{budget}.
Each individual in the study needs to be compensated for their participation, so the analyst
can only afford studies of limited size. This gives us the ingredients for a model for the analyst.
\begin{definition}
  The \emph{analyst} runs a private mechanism $M$ parameterized by $\epsilon$
  and $N$.  The mechanism comes with a real-valued \emph{accuracy
  function} $A_M (\epsilon, N)$, where smaller values of $A_M (\epsilon, N)$
  correspond to more accurate results. (We will omit the subscript when the
  mechanism is clear.) The analyst wants a target accuracy $\alpha$, and so
  requires that $A_M(\epsilon, N) \leq \alpha$.  Finally, the analyst has a
  \emph{budget} $B$ to spend on the study.
\end{definition}
Depending on what the analyst is trying to learn, he may be able to tolerate a
lower or higher total error.\iffoot\footnote{For an extreme case, suppose the
  analyst wants to learn a target value $v$ in the interval $(0, 1)$ (say, a
  fraction), and $A_M(\epsilon, N)$ is how far the private estimate $p(v)$ is
  from $v$, i.e., $A_M(\epsilon, N) = |p(v) - v|$.  Then,
  \bcp{Still don't understand this.}%
  \jh{Clearer?}%
values of $\epsilon, N$ such that $A_M(\epsilon, N) \geq 1$
do not promise useful accuracy---$p(v)$ may give no information about $v$.
}\fi{}
In general, the analyst may have a utility
function that quantifies how bad a specific amount of error is. Though our model
can be extended to handle this situation, for simplicity we assume that
the analyst cannot tolerate inaccuracy beyond the target level and is
equally happy with any inaccuracy within this level.

\SUBSECTION{The individual's view}
We next consider the individuals who might want to contribute their information
to a database in exchange for payment.
Study participants may want compensation for various reasons; for example, they
may want a flat compensation just for their time. Even though our model can be
extended to handle per-participant costs, for simplicity we do not consider this
cost.
Instead, we focus on the compensation most relevant to privacy:
participants may face personal harm if their data is revealed, so they are
willing to join the study only if they are adequately compensated for the risk
they take. A simple way to model the individual's risk is via a cost function
$f$, as described in Section~\ref{sec:background}.

We suppose the individual is offered a choice between participating in a
study and declining, but the study will always take place. Our model
does not say whether to run the study or not---are the study's potential
discoveries worth the potential (non-privacy-related) harm to
individuals?\footnote{Indeed, the difference in harm between running a
  study and not running the study may be very large: for instance, running a
  study may discover that smoking causes lung cancer, increasing costs for all
  smokers. }{}
Instead, we assume that some authority has decided that the study will take
place, and the individual only gets to decide whether to participate or not.
Thus, the individual participates only if they are compensated for their {\em
  marginal} increase in expected cost.

From the interpretation of differential privacy in terms of bad events
(Section~\ref{sec:interp}), an individual's expected cost should increase by at
most an $e^\epsilon$ if she decides to participate in a study. There is one
detail we need to attend to: our previous calculation of the marginal increase
in cost depends on the probability of each possible output of the mechanism.
This probability should be interpreted as taken over not only the randomness in
the mechanism, but also over the uncertainty of an individual about the rest of
the database.

To make this point clearer, we separate these two sources of randomness in the
calculation of the marginal increase in cost for a specific individual $x$. Let
$\mathcal{D}$ be the set of all possible databases containing $x$'s record, and
let $E$ be $x$'s expected cost if she decides {\em not} to participate.
Unpacking,
\begin{align*}
E &= \mathbb{E} [ f (M(D)) ] = \sum_{s \in \mathcal{R}, D^* \in
  \mathcal{D}} \Pr[ D = D^*, s = M(D) ] \cdot f(s) \\
 &= \sum_{D^* \in \mathcal{D}} \Pr[D = D^*] \cdot \sum_{s \in
   \mathcal{R}} \Pr[ s = M(D) \mid D = D^* ] \cdot f(s),
\end{align*}
where $\Pr[D = D^*]$ encodes an individual's belief about the contents of the
entire database, and by extension an individual's belief about the output of the
mechanism run on the entire database.
$E$ represents an upper bound on the individuals' beliefs about how much the
study will cost them if they do {\em not} participate in the study.  For
example, a study might discover that people in a certain town are likely to have
cancer---this knowledge could harm all the residents of the town, not just the
participants.
%
Similarly, if $C$ is the individual's expected cost if they {\em do} participate and $y$ is
any record in $D$ (representing a default or dummy record),
%
\begin{align*}
  C &= \mathbb{E} [ f (M(D \cup x \setminus y)) ] \\
  &= \sum_{s \in \mathcal{R}, D^* \in
  \mathcal{D}} \Pr[ D = D^*, s = M(D \cup x \setminus y) ] \cdot f(s) \\
  &= \sum_{D^* \in \mathcal{D}} \Pr[D = D^*] \cdot \sum_{s \in \mathcal{R}}
  \Pr[ s = M(D \cup x \setminus y) \mid D = D^* ] \cdot f(s).
\end{align*}
But the inner summation is the individual's expected cost when the rest of
the database is known to be $D^*$. By Equation~\EQREF{eq:dp-costs}, we
bound the increase of cost $C$ if $x$ participates (for any $y$):
\iffull
  \begin{align*}
    \sum_{s \in \mathcal{R}} \Pr[ s = M(D \cup x \setminus y) \mid D = D^* ]
    \cdot f(s)
    \sleq
    e^\epsilon \sum_{s \in \mathcal{R}} \Pr[ s = M(D) \mid D = D^* ] \cdot f(s).
  \end{align*}
\fi
Repacking the expressions for $E$ and $C$, we get $C \leq e^\epsilon E$. Hence
the individual's {\em marginal} cost of participation $C - E $ satisfies  $C - E
\leq e^\epsilon E - E = (e^\epsilon - 1) E$.


Now, we are ready to define a model for the individual.


\begin{definition}
  The \emph{individuals} are offered a chance to participate in a study with a
  set level of $\epsilon$ for some payment. Each individual considers a space of
  real-world events that, conditioned on the output of the study and the
  database size, are independent of their participation.

  Each individual also has a non-negative cost function on this space, which
  gives rise to a non-negative cost function $f$ on the space of outputs of the
  mechanism, and \emph{base cost} $\mathbb{E} [ f( R ) ]$,
  where $R$ is the random output of the mechanism
  without the individual's data. Let $E$ be an upper bound on the individual's
  base costs.
The individual participates only if they are compensated for the worst-case
increase in their expected cost by participating: $(e^\epsilon - 1) E$.
\end{definition}

Note the requirement on the space of bad events: we condition on
the output of the mechanism, as well as the size of the database. Intuitively,
this is because the size of the database is usually published. While such
information may sometimes be private,\iffoot\footnote{%
  In the worst case, an adversary may know the exact count of individuals with
  some disease to within $1$, in which case publishing the number of individuals with the
  disease could violate an individual's privacy.
}\fi{}
it is hard to imagine conducting a study without anyone knowing how many people
are in it---for one thing, the size controls the budget for a study. By this
conditioning, we require that an adversary cannot infer an individual's
participation even if he knows both the database size and the output of the
mechanism.

\SUBSECTION{Combining the two views}
To integrate the two views, we assume that the analyst directly compensates the
participants. Suppose the analyst has total budget $B$; since $N$ individuals
need to be paid $(e^\epsilon - 1) E$ each, we have the following budget
constraint:\footnote{Since we do not consider compensating participants for
  their time (though our model can be extended to cover this case), the
  ``budget'' should be thought of as the cost needed to cover privacy-related
  harm, part of a potentially larger budget needed to conduct the study. }

\begin{equation} \label{eq:budget}
  (e^\epsilon - 1) E N \sleq B
\end{equation}
%
This constraint, combined with the analyst's accuracy constraint
$A_{M}(\epsilon,N)\leq \alpha$, determines the feasible values of $N$ and
$\epsilon$. In general, there may not be any feasible values: in this case,
the mechanism cannot meet the requirements of the study. On the other
hand, there may be multiple feasible values. These trade off between the
analyst's priorities and the individual's priorities: larger values of
$\epsilon$ and smaller values of $N$ make the study smaller\iffoot\footnote{%
  In reality, the cost of the study also scales according to the size. It is not
  difficult to incorporate this into our model, but for simplicity we leave it
  out. Also, there may be a hard cap on the possible size of the study; we
  consider this situation in Section~\ref{sec:irqdb}.}\fi{}
and more accurate, while smaller values of $\epsilon$ and larger values of $N$
give a stronger guarantee to the individuals. In any case, feasible values of
$N$ and $\epsilon$ will give a study that is under budget, achieves the target
accuracy, and compensates each individual adequately for their risk. 

Note that the payments  depend on the particular study only through the $E$
parameter---different studies require different data, which may lead to
different base costs---and the $\epsilon$ parameter, which controls the privacy
guarantee; other internal details about the study do not play a role in this
model.
By using differential privacy as an abstraction, the model automatically covers
differentially private mechanisms in many settings: offline, interactive,
distributed, centralized, and more.  Further, the model can be applied whether
the analyst has benevolent intentions (like conducting a study) or malicious ones
(like violating someone's privacy).  Since differential privacy does not make
this distinction, neither does our model.

\SUBSECTION{Deriving the cost \texorpdfstring{$E$}{E}}
While the expected cost of not participating in a study may seem like a simple
idea, there is more to it than meets the eye. For instance, the cost may depend
on what the individuals believe about the outcome of the study, as well as what
bad events individuals are worried about. The cost could even depend on prior
private studies an individual has participated in---the more studies, the higher
the base cost.

Since individuals have potentially different beliefs about this cost, the
analyst must be sure to offer enough payment to cover each individual's expected
cost.  Otherwise, there could be {\em sampling bias}: individuals with high cost
could decline to participate in the study. While the analyst would like to offer
each individual just enough compensation to incentivize them to participate,
this amount may depend on private data. Thus, we model the analyst as paying
each individual the same amount, based on some maximum cost $E$.

Even if this maximum expected cost is difficult
to perfectly calculate in practice, it can be estimated in various ways:
reasoning about specific bad events and their costs, conducting surveys, etc.
While there has been work on using auctions to discover
costs related to privacy \cite{GR11,LR12,AH12,DFI12,RS12}, estimating this cost
in a principled way is an area of current research. Therefore, we will not pick
a single value of $E$ for our examples; rather, we show how different values
of $E$ affect our conclusions by considering estimates for a few scenarios.

\begin{remark} \label{rem:big}
Our goal in the following sections is to demonstrate how our model works in a
simple setting; as such, we will consider studies with very primitive
statistical analyses. As a result, the number of participants (and costs)
required to achieve a given level of accuracy may be unrealistically high.
There is a vast literature on sophisticated study design; more advanced methods
(such as those underlying real medical studies) can achieve better accuracy for
far less resources.
\end{remark}

\section{A simple study} \label{sec:ex-simple}

In this section, we will show how to apply our model to a simple study that
answers some queries about the database.

\SUBSECTION{A basic example: estimating the mean}

Suppose we are the analyst, and we want to run a study estimating the proportion
of individuals in the general population with some property $P$; we say this
target proportion $\mu$ is the \emph{population mean}. We also have a  measure
of accuracy $A(\epsilon, N)$ (which we define below), a target accuracy
level $\alpha$
and a fixed budget provided by the funding agency.

First, we specify our study.
For any given $N$ and $\epsilon$, we will recruit $N$ subjects to form a private
database $D_N$. We model the participants as being chosen independently and
uniformly at random, and we
consider the database $D_N$ as a random variable. (We sometimes call the
database the {\em sample}.) We then calculate the proportion of participants
with property $P$ (the \emph{sample mean})---call it $g(D_N)$. Since
$g$ is a $1/N$-sensitive function, we release it using the $\epsilon$-private
Laplace mechanism by adding noise $\nu (\epsilon, N)$ drawn from $Lap(1/N
\epsilon)$ to $g(D_N)$. 

Now, we can specify the accuracy function $A(\epsilon,N)$ of this study.  
%
In general, there are several choices of what $A$ can measure.  In this example,
we will fix the desired error $T$ and let $A$ be the probability of exceeding
this error.  Here, we consider the deviation from the
true population mean $\mu$ as our error.  We
say the mechanism \emph{fails} if it exceeds the error guarantee $T$, so $A$ is
the \emph{failure probability}.  Thus, we define
\[
  A(\epsilon, N) := \Pr[\, |g(D_N) + \nu(\epsilon, N) - \mu| \geq T ].
\]

There are two sources of error for our mechanism: from the sample mean deviating
from the population mean ($|g(D_N) - \mu|$), and from the Laplace noise added to
protect privacy ($|\nu(\epsilon, N)|$). Let us bound the first source of error.
\begin{theorem}[Chernoff Bound]
  \label{thm:chernoff}
  Suppose $\{ X_i \}$ is a set of $N$ independent, identically distributed $0/1$
  random variables with mean $\mu$ and sample mean $Y = \frac{1}{N} \sum X_i$.
  For $T \in [0, 1]$,
\[
  \Pr[\, |Y - \mu| \geq T ] \sleq 2 \exp \left( \frac{-N T^2}{ 3 \mu} \right).
\]
\end{theorem}

Assuming that our sample $D_N$ is drawn independently from the population,
we can model each individual as a random variable $X_i$ which is $1$ with
probability $\mu$, and $0$ otherwise. Then, the Chernoff bound is a bound on the
probability of the sample mean $g(D_N)$ deviating from the true proportion
$\mu$ that we are interested in.
Note that the sample must be free of sampling bias for this to hold---inferring
population statistics from a non-representative sample will skew the estimates.
This is why we must compensate participants so that they are incentivized to
participate, regardless of their private data.

Similarly, we use the following result to bound the second source of error, from
adding Laplace noise.

\begin{theorem}[Tail bound on Laplace distribution]
  \label{thm:lap-tail}
  Let $\nu$ be drawn from $Lap(\rho)$. Then,
\[
    \Pr [\, |\nu| \geq T ] \sleq \exp \left( - \frac{T}{\rho} \right).
\]
\end{theorem}

Now, since the total error of the mechanism is the difference between the sample
mean and the population mean plus the Laplace noise, if the output of the
Laplace mechanism deviates from the population mean by at least $T$, then either
the sample mean deviates by at least $T/2$, or the Laplace noise added is of
magnitude at least $T/2$.
Therefore, we can bound the failure probability $A(\epsilon,
N)$ by $$\Pr[\, |g(D_N) - \mu| \geq T/2 ] + \Pr [\, |\nu| \geq T/2 ].$$  Consider the
first term. Since $\mu \leq 1$, the Chernoff bound gives
\[
  \Pr [\, |g(D_N) - \mu| \geq T/2]  \sleq 2 e^{-N T^2 / 12 \mu} \sleq 2 e^{-N
    T^2 / 12}.
\]
The tail bound on the Laplace distribution gives
\[
  \Pr [\, |\nu(\epsilon, N)| \geq T/2] \sleq \exp \left( - \frac{T N
  \epsilon}{2} \right),
\]
since we added noise with scale $\rho=1/\epsilon N$. Therefore, given
a target accuracy $\alpha$, the accuracy constraint is
\begin{equation} \label{eq:medical-acc}
  A(\epsilon, N) := 2 \exp \left(- \frac{NT^2}{12} \right) + \exp \left( -
  \frac{T N \epsilon}{2} \right) \sleq \alpha.
\end{equation}
For the budget side of the problem, we need to compensate each individual by
$(e^\epsilon - 1)E$, according to our individual model. If our budget is $B$,
the budget constraint is Equation~\EQREF{eq:budget}:
\begin{mathdisplayfull}
  (e^\epsilon - 1 ) E N \sleq B.
\end{mathdisplayfull}%

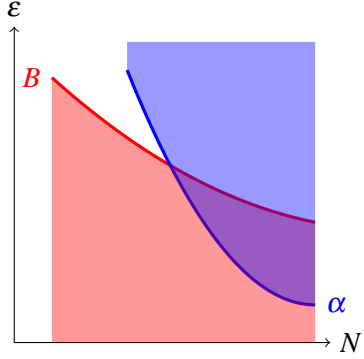
\begin{figure}[t]
  \centering
  \begin{tikzpicture}
    \draw[<->] (0,4.2) node[above] {$\epsilon$} -- (0,0) -- (4.2,0) node[right] {$N$};
    \draw[very thick, color=red] plot[domain=4:0.5] (\x, {0.1*\x*\x - 1*\x + 4})
    node[left] {$B$};
    \draw[very thick, color=blue] plot[domain=1.5:4] (\x, {0.5*\x*\x - 4*\x +
    8.5}) node[right] {$\alpha$};
    \fill[color=red, fill opacity=0.4] (4,0) -- plot[domain=4:0.5] (\x,
    {0.1*\x*\x - 1*\x + 4}) -- (0.5,0);
    \fill[color=blue, fill opacity=0.4] (1.5,4) -- plot[domain=1.5:4] (\x, {0.5*\x*\x - 4*\x +
    8.5}) -- (4,4);
  \end{tikzpicture}
  \caption{Feasible $\epsilon, N$, for accuracy $\alpha$ and budget $B$.}
  \label{fig:eps-N}
\end{figure}

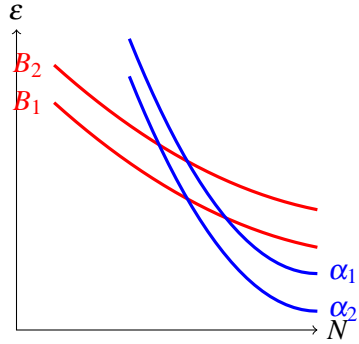
\begin{figure}[t]
  \centering
  \begin{tikzpicture}
    \draw[<->] (0,4) node[above] {$\epsilon$} -- (0,0) -- (4,0) node[right] {$N$};
    \draw[very thick, color=red] plot[domain=4:0.5] (\x, {0.1*\x*\x - 1*\x + 3.5})
    node[left] {$B_1$};
    \draw[very thick, color=red] plot[domain=4:0.5] (\x, {0.1*\x*\x - 1*\x + 4})
    node[left] {$B_2$};
    \draw[very thick, color=blue] plot[domain=1.5:4] (\x, {0.5*\x*\x - 4*\x +
    8.25}) node[right] {$\alpha_2$};
    \draw[very thick, color=blue] plot[domain=1.5:4] (\x, {0.5*\x*\x - 4*\x +
    8.75}) node[right] {$\alpha_1$};
  \end{tikzpicture}
  \caption{Constant accuracy curves for $\alpha_1 < \alpha_2$, constant budget
  curves for $B_1 < B_2$}
  \label{fig:eps-N-s}
\end{figure}

Our goal is to find $\epsilon$ and $N$ that satisfy this budget constraint, as
well as the accuracy constraint Equation~\EQREF{eq:medical-acc}. While it is
possible to use a numerical solver to find a solution, here we derive a
closed-form solution. Eliminating $\epsilon$ and $N$ from these
constraints is difficult,
so we find a sufficient condition on feasibility instead. First, for large
enough $\epsilon$, the sampling error dominates the error introduced by the
Laplace noise. That is, for $\epsilon \geq T/6$,

\[
  \exp \left( -\frac{T N \epsilon}{2} \right) \sleq \exp \left( -
  \frac{NT^2}{12} \right),
\]
%
so it suffices to satisfy this system instead:
\iffull
\begin{align}
  3 \exp \left( \frac{-N T^2}{12} \right) &\sleq \alpha \notag \\
  (e^\epsilon - 1) E N &\sleq B \label{eq:simple-constr}.
\end{align}
\fi
Figure~\ref{fig:eps-N} gives a pictorial representation of the constraints in
Equation~\EQREF{eq:simple-constr}. For
a fixed accuracy $\alpha$, the blue curve (marked $\alpha$) contains values of
$\epsilon, N$ that achieve error $\alpha$. The blue shaded region (above the
$\alpha$ curve) shows points that are feasible for that accuracy---there,
$\epsilon, N$ give accuracy better than $\alpha$. The red curve (marked $B$) and
red shaded region (below the $B$ curve) show the same thing for a fixed budget
$B$. The intersection of the two regions (the purple area) contains values of
$\epsilon, N$ that satisfy both the accuracy constraint, and the budget
constraint. Figure~\ref{fig:eps-N-s} shows the equality curves for
Equation~\EQREF{eq:simple-constr} at different fixed values of $\alpha$ and $B$.

Solving the constraints for $N$, we need
\begin{equation} \label{eq:study-n}
  N \sgeq \frac{12}{T^2} \ln \frac{3}{\alpha}.
\end{equation}
%
Taking equality gives the loosest condition on $\epsilon$, when the second
constraint becomes
%
\[
  \epsilon \sleq \ln \left( 1 + \frac{B T^2}{12 E \ln \frac{3}{\alpha}}
  \right).
\]
%
Thus, combining with the lower bound on $\epsilon$, if we have
\begin{equation}
  \frac{T}{6} \sleq \epsilon\sleq \ln \left( 1 + \frac{B T^2}{12 E \ln
    \frac{3}{\alpha}} \right),
  \label{eq:study-feas}
\end{equation}
then the study can be done at accuracy $\alpha$, budget $B$.
Since we have assumed $\epsilon \geq T/6$, this condition is sufficient but
not necessary for feasibility. That is, to deem a study {\em
  infeasible}, we need to check that the {\em original} accuracy constraint
(Equation~\EQREF{eq:medical-acc}) and budget constraint
(Equation~\EQREF{eq:budget}) have no solution.

For a concrete instance, suppose we want to estimate the true proportion with $\pm 0.05$
accuracy ($5\%$ additive error), so we take $T = 0.05$. We want this accuracy
except with at most $\alpha = 0.05$ probability, so that we are $95\%$ confident
of our result. If Equation~\EQREF{eq:study-feas} holds, then we can set
$\epsilon = T/6 = 0.0083$ and $N$ at equality in Equation~\EQREF{eq:study-n},
for $N \approx 20000$.

For the budget constraint, let the budget be $B = 3.0 \times 10^4$.
If we now solve Equation~\EQREF{eq:study-feas} for the base cost $E$, we
find that the study is feasible for $E \leq E_{feas} \approx 182$.
This condition also implies a bound on an individual's cost increase,
which we have seen is $(e^\epsilon - 1) E$. To decide whether a study is
feasible, let us now now estimate the cost $E$ and compare it to $E_{feas}$ in
various scenarios.

\SUBSECTION{Analyzing the costs scenarios}

We consider participant costs for four {\em cost scenarios}, which will serve as
our running examples throughout the paper.

\noindent{\bf Smoking habits.} 
The data comes from smokers who fear their health insurance company will find
out that they smoke and raise their premiums.
The average health insurance premium difference between  smokers and nonsmokers
is $\$1{,}274$~\cite{moriarty}.  Thus, some participants fear a price increase
of $\$1{,}274$.\iffoot\footnote{%
  Note that it would not make sense to include the participant's current
  health insurance cost as part of the bad event---participating in
  a study will not make it more likely that participants need to pay for health
  insurance (presumably, they are already paying for it). Rather, participating
  in a study may lead to a payment {\em increase}, which is the bad event.
}\fi{}
Since we pay the same amount to everyone, smoker or not, we base our estimate of
the base cost on the smokers' concerns.
To estimate the base cost $E$, we first estimate the probability that the
insurance company concludes that an individual smokes, even if they do not
participate in the study. This is not impossible: perhaps people see her
smoking outside.

So, suppose the participants think there is a moderate, $20\%$ chance
that the insurance company concludes that they are smokers, even if they do not
participate.\iffoot\footnote{%
  Nonsmokers would expect much lower cost, but not zero: there is always a
  chance that the insurance company wrongly labels them a smoker.
}\fi{}
Thus, we can estimate the base cost by $E = 0.20 \cdot 1274 = 254.8$---this is
the cost the participants expect, even if they do not participate. Since this is
far more than $E_{feas} \approx 182$, the study may not be feasible. To check,
we plug the exact accuracy constraint (Equation~\EQREF{eq:medical-acc}) and
budget constraint (Equation~\EQREF{eq:budget}) into a numeric solver, which
reports that the system has no solution for $\epsilon, N$. Thus, we conclude
that the study not feasible.

Perhaps this is not so surprising: since smoking is often done in public, it may
not be considered truly private information---indeed, the probability of the bad
event if the individual did not participate was already fairly large (we
estimated $20\%$).  For more typical privacy scenarios, this may not be the
case.

\noindent{\bf Educational data.}
The data consists of students' educational records, including courses taken and
grades.  Suppose individuals fear their record of classes and grades are
published, perhaps causing them to be fired and forced to switch to a job with
lower pay.  The mean starting salary of college graduates in the United States
is $\$45{,}000$~\cite{nace-survey}; let us suppose that they face a pay cut of
$30\%$ ($\$12{,}500$) if their records become public.

If the individual does not participate in the study, it is still possible that
an employer uncovers this data: perhaps someone steals the records and publishes
them, or maybe the outcome of the study (without the individual) can be combined
with public information to infer an individual's grades. However, complete
disclosure seems like a low probability event; let us suppose the chance is one
in a thousand ($0.001$). Then, we can estimate the base cost by $E = 0.01 \cdot 12500 =
12.5$. Since this is less than $E_{feas}$, the study is feasible.

\noindent{\bf Movie ratings data.} 
The data consists of movie ratings; suppose individuals fear their private movie
ratings are published, much like the case of the Netflix
challenge~\cite{narayanan-2008-deanonymization}.
Again, while an exact monetary amount of harm is not obvious, we can consider
the monetary cost of this disclosure, legally speaking. The Video Privacy
Protection Act of 1998\iffoot\footnote{%
  This law was enacted after Supreme Court nominee Robert Bork's video store
  rental records were written up in a newspaper.}\fi{}
is an American law stating that punitive damages for releasing video rental
records should be at least $\$2{,}500$.

If the individual does not participate in the study, it is possible that their
movie ratings are still published---perhaps their internet service provider is
monitoring their activities, or someone guesses their movies. Again, disclosure
seems like a low probability event; let us suppose the chance is one in
ten thousand ($0.00001$). Then, we can estimate the base cost by $E =
0.0001 \cdot 2500 = 0.25$. Since this is less than $E_{feas}$, the study is
feasible.

\noindent{\bf Social networks.}
Many social networks allow anonymous personas. Suppose individuals fear their
true identity will be revealed from studies on the social network structure,
much like deanonymization attacks against anonymous Twitter
accounts~\cite{NS-social}.
The exact monetary amount harm of disclosure is not clear, but the dangers can
be serious: consider dissidents operating anonymous accounts. Suppose we
estimate the cost of disclosure to be very high, say $\$100{,}000$.

If the individual's information is not included in the social network data, it
is unlikely that their identity is revealed---perhaps there is a physical privacy
breach. If the individual takes strong precautions to guard their privacy, then
disclosure seems like a very low probability event if they do not participate in
this study. Let us suppose the chance is one in a hundred thousand ($0.00001$). Then,
we can estimate the base cost by $E = 0.00001 \cdot 100000 = 1$. This is less
than $E_{feas}$, so the study is feasible.

\SUBSECTION{A more realistic example: answering many queries}

When working out the simple study above, we needed to derive the accuracy
totally from scratch.
Moreover, the mechanism is not very powerful---it can answer only a single
query! In this section, we address these issues by considering a more
sophisticated algorithm from the privacy literature: the multiplicative weights
exponential mechanism (MWEM)~\cite{HLM12,HR10}. As part of our analysis, we will
show how to directly plug established accuracy results for MWEM into our model.

MWEM is a mechanism that can answer a large number of {\em counting queries}:
queries of the form ``What fraction of the records in the database satisfy
property $P$?'' For example, suppose that the space of records is bit strings of
length $d$, i.e. $\mathcal{X} = \{0, 1\}^{d}$.  Each individual's bit string can
be thought of as a list of attributes: the first bit might encode the gender,
the second bit might encode the smoking status, the third bit might encode
whether the age is above $50$ or not, etc. Then, queries like ``What fraction of
subjects are male, smokers and above $50$?'', or ``What proportion of subjects
are female nonsmokers?'' are counting queries.

To use our model, we will need an accuracy bound for MWEM~\cite{HR10}: For a
data universe $\mathcal{X}$, set of queries $\mathcal{C}$ and $N$ records, the
$\epsilon$-private MWEM answers all queries in $\mathcal{C}$ within additive
error $T$ with probability at least $1 - \beta$, where
\begin{equation*}
  T = \left( \frac{128 \ln |\mathcal{X}| \ln
  \left( \frac{32 |\mathcal{C}| \ln |\mathcal{X}|}{\beta T^2}
\right)}{\epsilon N} \right)^{1/3}.
\end{equation*}

We again define the accuracy function $A(\epsilon, N)$ to be the probability
$\beta$ of exceeding error $T$ on any query.  Solving,
\[
  A(\epsilon, N) := \beta = \frac{32 |\mathcal{C}| \ln |\mathcal{X}|}{T^2} \exp
  \left( -\frac{\epsilon N T^3}{128 \ln |\mathcal{X}|} \right),
\]
Like the previous example, we want to satisfy the constraints
$A(\epsilon, N) \leq \alpha$ and  $(e^\epsilon - 1)EN \leq B$.

For a concrete example, suppose we want $\mathcal{X} = \{0, 1\}^{8}$ so
$|\mathcal{X}| = 2^{8}$ and accuracy $T = 0.2$ for $20\%$ error, with bad
accuracy at most $5\%$ of the time, so $\alpha = 0.05$. Further, we want to
answer $10000$ queries,
so $|\mathcal{C}| = 10000$.

We can now carry out cost estimates; suppose that the
budget is $B = 2.0 \times 10^6$. Looking at the movie ratings scenario when $E =
0.25$, the constraints are satisfiable: take $N =8.7 \times 10^5$, $\epsilon =
2.3$, compensating each individual $(e^\epsilon - 1) E = 2.2$.

For the social network scenario where $E = 1$, the constraints are also
satisfiable: take $N = 1.3 \times 10^6$, $\epsilon = 1.5$, compensating each individual
$(e^\epsilon - 1) E = 3.5$. The calculations are similar for the other cost
scenarios; since those examples have higher base cost $E$, the required budget
will be higher.

\section{The true cost of privacy} \label{sec:cost-of-privacy}


Now that we have a method for estimating the cost of a private study, we can
compare this cost to that of an equivalent non-private study. Again, we consider
only costs arising from compensation for harm to privacy.

Differential
privacy requires additional noise to protect privacy, and requires a
larger sample to achieve the same accuracy.  Hence, one would expect private
studies to be more expensive than equivalent non-private studies.
While this is true if individuals are paid the same in both cases, differential
privacy also has an advantage: it can bound the harm to individuals,
whereas---as de\-a\-no\-ny\-mi\-za\-tion attacks have shown---it is
difficult for non-private studies to make any guarantees about privacy.

When an individual participates in a non-private study, it is very possible that
their information can be completely recovered from published results.
Thus, to calculate the cost for the non-private study, we consider a
hypothetical world where non-private study participants are compensated
in proportion to their worst-case cost $W$, i.e., their cost for having their
data published in the clear.\iffoot\footnote{Note the difference between $W$
  and $E$: the former measures the harm to the individual of revealing private
  data, while the latter measures the harm to the individual from running the
  study, even without the individual's participation.  Naturally, $W$ is much
  higher than $E$---the expected harm from having private data directly
  published is usually greater than the expected harm of not even contributing
  private information.

  While $E$ may include costs for the bad event ``an adversary guesses my data
  and publishes it in the clear,'' which would lead to cost $W$, this cost
  should be weighted by the (very low) probability of disclosure if the
  individual does not even contribute their data. }\fi{}

It would be unreasonable for the non-private study to pay each
individual their full worst-case cost: even in the most poorly designed non-private
study, it is hard to imagine every individual having their data published in the
clear. More likely, attackers may be able to recover some fraction of the data;
for instance, if attackers are analyzing correlations with public datasets, the
public datasets may contain information for only a portion of all the study
participants. Thus, we suppose a non-private study might expose up to a $\phi$
fraction of the participants, and the non-private study compensates each
participant with a $\phi$ fraction of their worst-case cost, i.e., $\phi W$.

Consider the mean-estimation study from Section~\ref{sec:ex-simple}.  For the
non-private study with $N'$ individuals, we directly release the sample mean
$g(D) = \frac{1}{N'} \sum X_i$.\footnote{%
  Remark~\ref{rem:big} holds here as well: more sophisticated statistical
  experiments can achieve accuracy for less resources, while we use a simple
  analysis. However, we do so for {\em both} the private and non-private
  studies, so comparing the relative costs is fair.}
Thus, we do not have to bound the error from Laplace noise---all the error is
due to the sample mean deviating from the population mean.

\iffull
First, we calculate the sample size $N'$ a non-private study needs in order to
achieve the same level of accuracy as the private study. This will determine the
minimum budget that is needed for the non-private study. We use the following
bound.

\begin{theorem}[Chernoff Lower Bound]
  \label{thm:chernoff-lower}
  Suppose  $\{ X_i \}$ are $N$ independent, identically distributed
  $0/1$ random variables with mean $\mu \leq 1/4$ and sample mean $Y =
  \frac{1}{N} \sum X_i$. For $T \in [0, 1]$,
\[
  \Pr[\, |Y - \mu| \geq T ] \sgeq \frac{1}{2} \exp \left( \frac{-2 N T^2}{\mu}
  \right).
\]
\end{theorem}
Intuitively, the standard Chernoff bound says that the probability of the sample
mean deviating from the population mean is \emph{at most} some value, but this
bound might be very loose: the true probability of deviation could be much lower.
The lower bound theorem says that the probability of deviation is also \emph{at
  least} some value. This will allow us to lower bound the number of individuals
needed to achieve a desired accuracy.
\fi

We can then give conditions under which the private study is cheaper than the
public study.

\begin{theorem} \label{thm:private-cheaper}
  Given a target error $T \geq 0$ and target accuracy $\alpha > 0$,  the private
  mean estimation study  will be cheaper than the non-private mean estimation
  study exposing a fraction $\phi$ of participants if the following (sufficient,
  but not necessary) condition holds:
  \begin{equation} \label{eq:private-cheaper}
    \frac{T}{6} \sleq \ln \left( 1 + \frac{\phi W \ln \frac{1}{2 \alpha}}{96 E
        \ln \frac{3}{\alpha}} \right)
  \end{equation}
  The non-private study needs at least $N'$ individuals, where
  \begin{equation} \label{eq:nonprivate-n}
    N' \sgeq \frac{1}{8 T^2} \ln \frac{1}{2 \alpha}.
  \end{equation}
\end{theorem}
\iffull
\begin{proof}
  First, we derive a lower bound on the sample size $N'$ that is necessary for the
  non-private study to have low error. Since we want the deviation probability to
  be at most $\alpha$ for all $\mu$, in particular the sample size must be large
  enough guarantee this error for $\mu = 1/4$. Theorem~\ref{thm:chernoff-lower}
  gives
  \[
    \alpha \sgeq \Pr[\, |Y - \mu| \geq T ] \sgeq  e^{-8 N' T^2}/2,
  \]
  which is equivalent to
  \[
    N' \sgeq \frac{1}{8 T^2} \ln \frac{1}{2 \alpha},
  \]
  as desired.

  Thus, the minimum budget for the non-private study is at least $B' = \phi N' W =
  \frac{\phi W}{8 T^2} \ln \frac{1}{2 \alpha}$, so by Equation~\EQREF{eq:study-feas},
  the private study will be cheaper than the non-private study if
  \[
    \frac{T}{6} \sleq \ln \left( 1 + \frac{\phi W \ln \frac{1}{2 \alpha}}{96 E
        \ln \frac{3}{\alpha}} \right).
    \vspace*{-3ex} 
  \]
\end{proof}

Recall that since Equation~\EQREF{eq:study-feas} is a sufficient but not
necessary condition on the private study being feasible, if the above equation
does not hold, we cannot conclude that the private study is necessarily more
expensive than the public one from Theorem~\ref{thm:private-cheaper}.
\fi

For our example calculations below, we take $\phi = 1/500 = 0.002$.\footnote{%
Precise measurements of the success
rate of real deanonymization attacks are hard to come by, for at least two
reasons: first, published deanonymization attacks aim to prove a concept, rather
than violate as many people's privacy as possible.  Second, published
deanonymization attacks generally do not have the luxury of knowing the original
data, so they are necessarily conservative in reporting success rates.
Adversarial attacks on privacy need not satisfy these constraints.
\jh{Netflix, they got two out of $500{,}000$ definitely right. But they were
  quite conservative, these two were like $15$ standard deviations away from
  normal. They got more around two standard deviations away, which is still
  fairly likely of being correct.}%
}{}
Now that we have a sufficient condition for when the private study is cheaper
than the public study, let us turn back to our four cost scenarios. 
From Section~\ref{sec:ex-simple}, recall we wanted to estimate the mean of a
population to accuracy $T = 0.05$, with failure probability at most $\alpha =
0.05$.

\noindent{\bf Smoking habits.}
Recall that we estimated the base cost $E$ to be $254.8$.  The worst case cost
is at least the rise in health insurance premium, so we let $W = 1274$. Plugging
in these numbers, Equation~\EQREF{eq:private-cheaper} does not hold. So, the
private study is not necessarily cheaper.

\noindent{\bf Educational data.}
Recall that we estimated the base cost $E$ to be $12.5$.  The worst case cost is
at least the loss in salary: $\$12{,}500$; we take this to be $W$. Plugging
these numbers into Equation~\EQREF{eq:private-cheaper}, we find that the private
study is cheaper.

\noindent{\bf Movie ratings.}
Recall that we estimated the base cost $E$ to be $0.25$, and we estimated the
worst case disclosure cost to be at least the damages awarded
under the Video Privacy Protection Act.  So we let $W = 2500$. Plugging these
numbers into Equation~\EQREF{eq:private-cheaper}, we find that the private study
is cheaper.

\noindent{\bf Social networks.}
Recall that we estimated the base cost $E$ to be $1$. The worst case cost is at
least the cost of discovery: $\$100{,}000$; we let this
be $W$.  Plugging these numbers into Equation~\EQREF{eq:private-cheaper}, we
find that the private study is cheaper.


Let us compare the size and costs of the two studies, just for the movie ratings
scenario.  By Equation~\EQREF{eq:nonprivate-n}, the non-private study needs $N'
\geq 115$. As expected, the non-private study needs fewer people to achieve
the same accuracy compared to the private study ($N = 20000$), since no noise is
added.  However, the total cost for non-private study would be $B' = \phi W N =
0.002 \cdot 2500 \cdot 115 \approx 575$. The equivalent private study, with $E =
0.25, \epsilon = 0.0083, N = 20000$ costs $(e^\epsilon - 1) \cdot EN \approx
40$.

If both private and non-private studies have the same budget, the private study
can buy more participants to further improve its accuracy. Thus, this private
study is {\em more accurate} and {\em cheaper} (and more private!) than the
non-private version.

\section{Extending the model}
So far, we have considered just two constraints on $\epsilon$ and $N$: expected
cost to the individuals (expressed as a budget constraint), and accuracy for the analyst. Other
constraints may be needed to model finer details---we will refer to these
additional constraints as {\em side conditions}. In this section, we first
consider generic upper and lower bounds on $\epsilon$---these follow from the
definition of differential privacy. Then, we present a case study incorporating
side conditions.

\SUBSECTION{Upper bounds on \texorpdfstring{$\epsilon$}{epsilon}}
While the definition of differential privacy is formally valid for any value of
$\epsilon$ \cite{DMNS11}, values that are too large or too small give weak
guarantees.  For large values of $\epsilon$, the upper bound on the probability
$\Pr[ M(D) \in S]$ can rise above one and thus become meaningless: for instance,
if $\epsilon = 20$, Equation~\EQREF{eq:dp-ubound} imposes no constraint on the
mechanism's output distribution unless $\Pr[ M(D') \in S] \leq
e^{-20}$.\iffoot\footnote{%
  Even though the upper bound may not guarantee anything, differential
  privacy still gives some guarantee. For instance, suppose $\epsilon = 20$ and
  $\Pr[ M(D') \in S ] = 1/2 > e^{-20}$. The upper bound gives
\[
  \Pr[ M(D) \in S ] \sleq e^{20} \cdot 1/2 \approx 10^9,
\]
which is useless. However, consider the outputs $\bar{S} = \mathcal{R}
\setminus S$: we know that $\Pr [ M(D') \in \bar{S} ] = 1/2$, so by
Equation~\EQREF{eq:dp-lbound},
\[
  \Pr[ M(D) \in \bar{S} ] \sgeq e^{-20} \cdot 1/2 \approx 10^{-9},
\]
which is a nontrivial bound. In particular, it implies that
\[
  \Pr[ M(D) \in S ] = 1 - \Pr [ M(D) \in \bar{S} ] \lesssim 1 - 10^{-9}.
\]
}\fi{}

To demonstrate this problem, we describe an $\epsilon$-private mechanism for
large $\epsilon$ which is not intuitively private.
Consider a mechanism $M$ with range $\mathcal{R}$ equal to data universe
$\mathcal{X}$, and consider a targeted individual $J$.  When $J$ is in the
database, the mechanism publishes their private record with probability $p^* >
1/|\mathcal{X}|$, otherwise it releases a record at random. 

We first show that this mechanism is $\epsilon$-differentially private, for
a very large $\epsilon$.
Let $j$ be $J$'s record, and let
\[
  p = \frac{1 - p^*}{|\mathcal{X}| - 1} < \frac{1}{|\mathcal{X}|}
\]
be the probability of releasing a record $s \neq j$ when $J$ is in the
database. Consider two databases $D \cup i$ and $D \cup j$, where $i$ is any
record.  For $M$ to be $\epsilon$-differentially private, it suffices that
\begin{align*}
  e^{-\epsilon} \Pr[ M(D \cup i) = j ] &\sleq \Pr[ M(D \cup j) = j ] \sleq e^\epsilon \Pr
  [ M(D \cup i) = j ] \\
  e^{-\epsilon} \Pr[ M(D \cup i) = s ] &\sleq \Pr[ M(D \cup j) = s ] \sleq e^\epsilon \Pr
  [ M(D \cup i) = s ],
\end{align*}
for all $s \neq j$. Rewriting, this means
\begin{equation*}
  e^{-\epsilon}\frac{1}{|\mathcal{X}|} \sleq p^* \sleq e^\epsilon
  \frac{1}{|\mathcal{X}|} \mbox{\quad and \quad}
  e^{-\epsilon} \frac{1}{|\mathcal{X}|} \sleq p \sleq e^\epsilon
  \frac{1}{|\mathcal{X}|}.
\end{equation*}
By assumption, the left inequality in the first constraint and the right
inequality in the second constraint hold. Thus, if
\begin{equation}
  \label{eq:eps-big-1}
  \epsilon \sgeq \ln (p^* |\mathcal{X}|),
\end{equation}
the first constraint is satisfied. Since the probabilities over all outputs sums
to one, we also know $p^* + (|\mathcal{X}| - 1) p = 1$. So,
\begin{equation}
  \label{eq:eps-big-2}
  \epsilon \sgeq \ln \left( \frac{1}{p |\mathcal{X}|} \right)
  \sgeq \ln \left( \frac{|\mathcal{X}| - 1}{|\mathcal{X}| (1 - p^*)} \right)
\end{equation}
suffices to satisfy the second constraint.

Therefore, $M$ is $\epsilon$-differentially private if $\epsilon$ satisfies
these equations.  For instance, suppose $|\mathcal{X}| = 10^6$, and $p^* =
0.99$. $M$ almost always publishes $J$'s record (probability
$0.99$) if $J$ is in the database, but it is still $\epsilon$-differentially
private if $\epsilon \geq 14$.

Clearly, a process that always publishes a targeted individual's data if they
are in
the database and never publishes their data if they are not in the database is
blatantly non-private. This $\epsilon$-private mechanism does nearly the same
thing: with probability $p^* = 0.99$, it publishes $J$'s record with
probability at least $p^* = 0.99$ if $J$ is in the database, and with probability
$1/|\mathcal{X}| = 10^{-6}$ if $J$ is not. Evidently, values of
$\epsilon$ large enough to satisfy both Equations~\EQREF{eq:eps-big-1} and
\EQREF{eq:eps-big-2} do not give a very useful privacy guarantee.

\SUBSECTION{Lower bounds on \texorpdfstring{$\epsilon$}{epsilon}}
While choosing $\epsilon$ too large will compromise the privacy guarantee,
choosing $\epsilon$ too small will ruin {\em accuracy}---the
mechanism must behave too similarly for databases that are very
different.
For example, let $D, D'$ be arbitrary databases of size $N$, and
let $0 < \epsilon \leq \frac{1}{N}$. Since the two databases have the same
size, we can change $D$ to $D'$ by changing at most $N$ rows. Call
the sequence of intermediate neighboring databases $D_1,\cdots,D_{N - 1}$. By
differential privacy,
\begin{align*}
  \Pr [ M(D) \in S ] &\sleq e^\epsilon \Pr [M(D_1) \in S] \\
  \Pr [ M(D_1) \in S ] &\sleq e^\epsilon \Pr [M(D_2) \in S] \\
\iffull &\cdots \\ \fi
  \Pr [ M(D_{N - 1}) \in S ] &\sleq e^\epsilon \Pr [M(D') \in S].
\end{align*}
Combining, $\Pr[M(D) \in S] \leq e^{N \epsilon} \Pr[M(D') \in S]$. Similarly, we
can use Equation~\EQREF{eq:dp-lbound} to show $\Pr[M(D) \in S] \geq e^{-N
  \epsilon} \Pr[M(D') \in S]$.  But we have taken $\epsilon \leq 1/N$, so the
exponents are at most $1$ and at least $-1$.  So, the probability of every event
is fixed up to a multiplicative factor of at most $e$, whether the input is $D$
or $D'$.  (Differential privacy with $\epsilon = 1$ guarantees this for
neighboring databases, but here $D$ and $D'$ may differ in many---or
all!---rows.) Such an algorithm is probably useless: its output distribution
depends only weakly on its input.\iffoot\footnote{%
  For an extreme case, a $0$-private mechanism has the same probability of
  releasing $s$ whether the database is $D$ or $D'$. Since this holds for all
  pairs of neighboring inputs, a $0$-private mechanism is useless---it behaves
  independently of its input.
}\fi{}

\SUBSECTION{Case Study: Educational statistics}\label{sec:irqdb}

Putting everything together, we now work through an example with these added
constraints on $\epsilon$, together with a limit on the study size. We consider the
same mean estimation study from Section~\ref{sec:ex-simple}, except now with
side constraints.


Concretely, suppose that we are in the educational data scenario, where
each student's record contains class year ($4$ possible values), grade point
average (rounded to the nearest letter grade, so $5$ possible values), years
declared in the major ($4$ possible values), and order of courses in the major
($100$ possible combinations). The total space of possible values is
$|\mathcal{X}| = 4 \cdot 5 \cdot 4 \cdot 100 = 8000$.

We now add in our side conditions. First, suppose that we have data for $N =
1000$ students, spanning several years. It may not be simple to expand the study
size---perhaps this data for all the students, or perhaps we only
have access to recent data. the only way to collect more data is to graduate more
students. We also want the upper and lower bounds on $\epsilon$ discussed above
to hold.

For the accuracy, recall from Section~\ref{sec:ex-simple} that if $T$ is the
desired additive error and $\alpha$ is the probability we do not achieve this
accuracy, the accuracy constraint is
\[
  2 \exp \left(- \frac{NT^2}{12} \right) + \exp \left( - \frac{T N \epsilon}{2}
  \right) \sleq \alpha.
\]
In this example it is not very natural to think of a total budget for compensation,
since all the data is assumed to have been already collected.\footnote{While our
  model assumes individuals have a choice to participate, it can be seen to
  apply even when individuals do not; for details, see
  Appendix~\ref{app:participate}.
}{}
Instead, we know the privacy harm for any individual is at most $(e^\epsilon -
1) \cdot E$, and we will bound the maximum allowed privacy harm per student.
Suppose it is $B_0 = 10$, giving the constraint
\begin{mathdisplayfull}
  (e^\epsilon - 1) \cdot E \sleq B_0.
\end{mathdisplayfull}

To capture the side conditions, we add a few more constraints. For the
population, we require $N \leq 1000$. For the upper bound on $\epsilon$, we do
not want Equations~\EQREF{eq:eps-big-1} and~\EQREF{eq:eps-big-2} to both hold, so we
add a constraint
\begin{equation*}
  \epsilon \sleq \max \left[ \ln (0.1 \cdot |\mathcal{X}|), \ln \left(
  \frac{|\mathcal{X}| - 1}{|\mathcal{X}| (1 - 0.1)} \right) \right].
\end{equation*}
For the lower bound on $\epsilon$, we add the constraint $\epsilon \geq 1/N$.

Putting it all together, with base cost $E = 12.5$, record space size
$|\mathcal{X}| = 8000$ and allowed harm per student $B_0 = 10$, and target error
$T = \alpha = 0.05$, the feasibility of this study is equivalent
to the following system of constraints.

\begin{mathpar}
  2 \exp \left(- 0.0002 \cdot N \right) + \exp \left( - 0.025 N \epsilon \right)
  \sleq 0.05, \and
  (e^\epsilon - 1) \cdot 12.5 \sleq 10, \and
  N \sleq 1000, \and
  1/N \sleq \epsilon \sleq \max (\ln (800), \ln \left( 1.11 \right))
\end{mathpar}
Note that we are requiring the same accuracy as in our original study in
Section~\ref{sec:ex-simple}, and in fact the original study without the side
constraints is feasible. However, a numeric solver shows that these constraints
are not feasible,
so this study is not feasible in this setting.\footnote{%
  To be precise, we have shown that this {\em particular} mechanism (i.e., the
  Laplace mechanism) is not feasible---there may be other, more clever
  mechanisms that feasibly compute what we want.  We are not aware of any such
  mechanisms, but we also cannot rule it out. }

\section{What about \texorpdfstring{$\delta$}{delta}?}\label{sec:delta}

In this section, we extend our model to an important generalization of
$\epsilon$-differential privacy, known as $(\epsilon, \delta)$-differential
privacy.

\begin{definition}[\cite{dwork2006our}]
\label{def:delta-priv}
  Given $\epsilon, \delta \geq 0$, a mechanism $M$ is \emph{$(\epsilon,
  \delta)$-differentially private} if for any two neighboring database $D, D'$,
  and for any subset $S \subseteq \mathcal{R}$ of outputs,
  \[
    \Pr [ M(D) \in S ] \sleq e^{\epsilon} \cdot \Pr [ M(D') \in S ] + \delta.
  \]
\end{definition}
Intuitively, this definition allows a $\delta$ probability of failure where the mechanism
may violate privacy. For instance, suppose that $s$ is an output that reveals
user $x$'s data. For a database $D'$ that does not contain user $x$'s
information, suppose $\Pr[M(D') \in S] = 0$. Under $\epsilon$-differential
privacy, $M$ can never output $s$ on any database. However, under
$(\epsilon,\delta)$-differential privacy, $M$ may output $s$ with
probability up to $\delta$, when fed any neighboring database $D$. In
particular if $D = D' \cup x \setminus y$, $\Pr[M(D) \in S]$ may be up to
$\delta$: even though $M$ never outputs $x$'s records on databases without $x$,
$M$ can output $x$'s record when she is in the database with probability
$\delta$.

\SUBSECTION{Modeling \texorpdfstring{$\delta$}{delta}}
By considering ``blatantly non-private'' mechanisms that nevertheless
satisfy $(\epsilon, \delta)$-privacy,
we can upper bound $\delta$. For example, for a database with $N$ records and
for $\delta = 1/N$, the mechanism that randomly outputs a record from the
database is $(0, \delta)$-private. This mechanism is intuitively non-private,
so we require $\delta \ll 1/N$ for a more reasonable guarantee.

For a more principled method of picking this parameter, we can model the costs
associated with different levels of $\delta$. The first step is to bound the
increase in expected cost for participating in an $(\epsilon, \delta)$-private
mechanism. We assume a bound $W$ on an individual's cost if their data is
publicly revealed, since with probability $\delta$ the mechanism may do just
that. Then, we can bound an individual's increase in expected cost when
participating in an $(\epsilon, \delta)$-private study.

\begin{proposition}
  Let $M$ be an $(\epsilon, \delta)$-private mechanism with range
  $\mathcal{R}$, and let $f$ be a non-negative cost function over $\mathcal{R}$.
  Let $W = \max_{s \in \mathcal{R}} f(s)$. Then, for neighboring databases $D,
  D'$,
%
\[
    \mathbb{E} [ f(M(D)) ] \sleq e^\epsilon \mathbb{E} [ f(M(D')) ] + \delta W.
\]
\end{proposition}
\iffull
\begin{proof}
  Let $\mathcal{R} = \{s_i\}$. For each output $s_i$,
\[
    \Pr[ M(D) = s_i ] = e^\epsilon \Pr[ M(D') = s_i ] + \delta_i,
\]
  where $\delta_i$ may be negative. Partition $\mathcal{R} = S^+ \cup S^-$,
  where $S^+$ contains the outputs $s_i$ with $\delta_i \geq 0$, and $S^-$
  contains the remainder. Now, from the definition of $(\epsilon,
  \delta)$-privacy, each $\delta_i$ is upper bounded by $\delta$, but we will
  show that not all $\delta_i$ can be $\delta$.
  Now,
  \begin{align*}
    \Pr[ M(D) \in S^+ ] &= \sum_{s_i \in S^+} \Pr [ M(D) = s_i ] \\
    &= \sum_{s_i \in S^+} e^\epsilon \Pr [ M(D') = s_i ] + \delta_i \\
    &= e^\epsilon \Pr [ M(D') \in S^+ ] + \sum_{s_i \in S^+} \delta_i.
  \end{align*}
  But from the definition of $(\epsilon, \delta)$-privacy applied to the set
  $S^+$, the left hand side is at most $e^\epsilon \Pr [ M(D') \in S^+ ] +
  \delta$. Hence,
%
\[
    \sum_{s_i \in S^+} \delta_i \sleq \delta.
\]
%
  Now we can conclude:
  \begin{align*}
    \mathbb{E} [ f(M(D)) ] &= \sum_{s_i \in \mathcal{R}} \Pr [ M(D) = s_i ]
    \cdot f(s_i) \\
    &\sleq \sum_{s_i \in \mathcal{R}} e^\epsilon \Pr [ M(D') = s_i ] \cdot
    f(s_i) + \delta_i \cdot f(s_i) \\
    &\sleq e^\epsilon \mathbb{E} [ f(M(D')) ] + \sum_{s^+_i \in S^+} \delta_i
    \cdot f(s^+_i) \\
    &\sleq e^\epsilon \mathbb{E} [ f(M(D')) ] + \delta W,
  \end{align*}
  as desired.
\end{proof}
\fi
We can now incorporate the $\delta$ parameter into our model.

\begin{definition}[$(\epsilon, \delta)$-private analyst model]
  An \emph{$(\epsilon, \delta)$-private analyst} is an
  analyst with accuracy $A_M$ a function of $\epsilon, N, \delta$.
\end{definition}

\begin{definition}[$(\epsilon, \delta)$-private individual model]
  An \emph{$(\epsilon, \delta)$-private individual} is an individual with a
  \emph{worst-case cost} $W$, which measures the cost of publicly revealing the
  individual's private information. The individual wants to be compensated for her
  worst-case marginal cost of participating under these assumptions:
  $e^\epsilon E + \delta W - E = (e^\epsilon - 1) E + \delta W$.
\end{definition}

Since $(\epsilon, \delta)$-privacy is weaker than pure $\epsilon$-privacy, why
is it a useful notion of privacy? It turns out that in many cases, $(\epsilon,
\delta)$-private algorithms are more accurate than their pure privacy
counterparts; let us consider such an example.

\SUBSECTION{Revisiting MWEM}
In Section~\ref{sec:ex-simple}, we analyzed the cost of MWEM.
We will now revisit that example with an $(\epsilon,\delta)$-private
version of MWEM. The setting remains the same: we wish to answer a large number
of counting queries with good accuracy, while preserving privacy.

The main difference is the accuracy guarantee, due to Hardt and
Rothblum~\cite{HR10}.
Suppose the space of records is $\mathcal{X}$ and 
we want to answer queries $\mathcal{C}$ to accuracy
$T$ with probability at least $1 - \beta$. The $(\epsilon,\delta)$-private MWEM
has accuracy
\begin{equation*}
  T = \frac{8 (\ln |\mathcal{X}| \ln(1/\delta))^{1/4} \ln^{1/2}
  \left( \frac{32 |\mathcal{C}| \ln |\mathcal{X}|}{\beta T^2} \right)}{N^{1/2}
  \epsilon^{1/2}}.
\end{equation*}
We define our accuracy measure $A(\epsilon, N)$ to be the failure probability
$\beta$. Solving, this means
\[
  A(\epsilon, N) := \beta = \frac{32 |\mathcal{C}| \ln |\mathcal{X}|}{T^2} \exp
  \left( - \frac{ \epsilon N T^2}{8 (\ln |\mathcal{X}| \ln (1/\delta))^{1/2}}
  \right).
\]
If $\alpha$ is the target accuracy, we need $A(\epsilon, N) \leq \alpha$.

For the budget constraint, we need $(e^\epsilon - 1) E N + \delta W N \leq B$.
Suppose we are in the social network scenario we described in
Section~\ref{sec:ex-simple}, with the same budget $B = 2.0 \times 10^6$ we used
for the $(\epsilon, 0)$-private MWEM algorithm. We use our running estimate of
the base cost for this scenario, $E = 1$, and the worst-case cost, $W = 10^6$.
For the other parameters, suppose the records are bit strings with $15$
attributes (versus $8$ before): $\mathcal{X} = \{0, 1\}^{15}$ and $|\mathcal{X}|
= 2^{15}$. We want to answer $|\mathcal{C}| = 200000$ queries (versus $10000$
before), to $5\%$ error (versus $20\%$ before), so $T = 0.05$, with probability
at least $95\%$ ($\alpha = 0.05$, same as before).

Plugging in the numbers, we find that the accuracy and budget constraints can
both be satisfied, for $\epsilon = 0.9$, $\delta = 10^{-8}$, and $N = 9.1 \times
10^5$. Each individual is compensated $(e^\epsilon - 1) E + \delta W = 1.46$,
for a total cost of $1.9 \times 10^6 \leq B$. Thus, the
$(\epsilon,\delta)$-private version of MWEM answers more queries, over a larger
space of records, to better accuracy, than the $(\epsilon, 0)$-version we
previously considered.


\section{Discussion}\label{sec:hard}

\SUBSECTION{Is all this complexity necessary?}

Compared to earlier threat models from the differential privacy literature, our
model may seem overly complex: the original definition from Dwork et
al.~\cite{dwork-2006-sensitivity} had only one parameter ($\epsilon$), while our model involves
a number of different parameters ($\alpha$, $A_M(\epsilon,N)$, $B$, and $E$).
So, at first glance, the original model seems
preferable. However, we argue that this complexity is present in the real world:
the individuals really \emph{do} have to consider the possible consequences of
participating in the study, the researchers really \emph{do} require a certain
accuracy, etc. The original definition blends these considerations into a single,
abstract number $\epsilon$. Our model is more detailed, makes the choices
explicit, and forces the user to think quantitatively about how
a private study would affect real events. 

\SUBSECTION{Possible refinements}

The key challenge in designing any model is to balance complexity and accuracy.
Our model is intended to produce reasonable suggestions for $\epsilon$ in most
situations while keeping only the essential parameters. Below, we review some
areas where our model could be refined or generalized.

\noindent\textbf{Estimating the base cost.}
Our model does not describe how to estimate the base cost for individuals. There
is no totally rigorous way to derive the base cost: this quantity depends on how individuals perceive
their privacy loss, and how individuals think about uncertain events.
These are both active areas of research---for instance, research in psychology has
identified a number of cognitive biases when people reason
about uncertain events~\cite{kahneman-1982-uncertainty}.  Thus, if the
consequences of participation are uncertain, the individuals might under- or
overestimate their expected cost.\footnote{ In some experiments, people give up
  their private data for as little as a
  dollar~\cite{beresford2012unwillingness}. }{}
More research is needed to incorporate these (and other) aspects of human
behavior into our model.

Another refinement would be to model individuals heterogeneously, with different
base costs and desired compensations. For instance, an individual who
has participated in many studies may be at greater risk than an individual who
has never participated in any studies. However, care must be taken to avoid
sampling bias when varying the level of compensation.

\noindent\textbf{Empirical attacks on privacy.}
Our model assumes that all $\epsilon$-private studies could potentially increase
the probability of bad events by a factor of $e^\epsilon$. It is not clear
whether (a) this is true for private algorithms considered in the literature,
and (b) whether this can be effectively and practically exploited. The field of
differential privacy (and our model) could benefit from empirical attacks on
private algorithms, to shed light on how harm actually depends on $\epsilon$,
much like parameters in cryptography are chosen to defend against known
attacks.

\noindent\textbf{Collusion.}
Our model assumes the study will happen regardless of a single individual's
choice. However, this may not be realistic if individuals collude: in an extreme case,
all individuals could collectively opt out, perhaps making a study impossible to
run.  While widespread collusion could be problematic, assumptions about the
size of limited coalitions could be incorporated into our model.

\noindent\textbf{Large $\epsilon$.}
As $\epsilon$ increases, our model predicts that the individual's marginal
expected harm increases endlessly. This is unreasonable---there should be
a maximum cost for participating in a study, perhaps the worst case cost
$W$. The cost curve could be refined for very small and very large values of
$\epsilon$.

\noindent\textbf{Modeling the cost of non-private studies.}
Our comparison of the cost of private and non-private studies uses a very crude
(and not very realistic) model of the cost of non-private studies. More
research into how much individuals want to be compensated for their private
data would give a better estimate of the true tradeoff between private and
non-private studies.



\section{Related work}\label{sec:relatedwork}

\iffull
\begin{table*}
  \centering
  \begin{tabular}{|l|l|l|}
    \hline
    Authors & Value(s) of $\epsilon$ & Application \\
    \hline
    \hline
    McSherry-Mahajan \cite{MM10} & $0.1$---$10$ & Network Trace Analysis \\
    Chaudhuri-Monteleoni \cite{CM08} & $0.1$ & Logistic Regression \\
    Machanavajjhala et al. \cite{MKAGV08} & $<7$  & Census Data \\
    Korolova et al. \cite{KKMN09} & $\ln 2, \ln 5, \ln 10$ & Click Counts \\
    Bhaskar et al. \cite{BLST10} & $1.4$  & Frequent Items \\
    Machanvajjhala et al. \cite{MKS11} & $0.5$---$3$  & Recommendation System \\
    Bonomi et al. \cite{BXCF12} & $0.01$---$10$ & Record Linkage \\
    Li et al. \cite{LQSC12} & $0.1$---$1$ & Frequent Items \\
    Ny-Pappas \cite{NP12} & $\ln 3$ & Kalman Filtering \\
    Chaudhuri et al. \cite{CSS12} & $0.1$---$2$ & Principal Component Analysis
    \\
    Narayan-Haeberlen. \cite{narayan-2012-djoin} & $0.69$ & Distributed Database Joins\\
    Chen et al. \cite{CRFG12} & $1.0 - 5.0$ & Queries over Distributed Clients\\
    Acs-Castelluccia \cite{AC12} & $1$ & Smart Electric Meters \\
    Uhler et al. \cite{USF12} & $0.1$---$0.4$ & Genome Data \\
    Xiao et al. \cite{XXFG12} & $0.05$---$1$ & Histograms \\
    Li-Miklau. \cite{LM12} & $0.1$---$2.5$ & Linear Queries \\
    Chen et al. \cite{CFD11} & $0.5$---$1.5$ & Trajectory Data \\
    Cormode et al. \cite{CPSSY11} & $0.1$---$1$ & Location Data \\
    Chaudhuri et al. \cite{CMS09} & $0.01$---$0.5$ & Empirical Risk
    Minimization \\
    \hline
  \end{tabular}
  \caption{Values of $\epsilon$ in the literature}
  \label{table:xyz}
\end{table*}
\fi

There is by now a vast literature on differential privacy, which we do not
attempt to survey. We direct the interested reader to an excellent survey by
Dwork \cite{dwork-2008-survey}.

The question of how to set $\epsilon$ has been present since the introduction of
differential privacy. Indeed, in early work on differential privacy, Dwork
\cite{dwork-2008-survey} indicates that the value of $\epsilon$, in economic
terms or otherwise, is a ``social question.'' Since then, few works have taken
an in-depth look at this question.  Works applying differential privacy have
used a variety of choices for $\epsilon$, mostly ranging from
$0.01$--$10$\iffull\ (see Table \ref{table:xyz})\fi, with little or no
convincing justification.

The most detailed discussion of $\epsilon$ we are aware of is due to Lee and
Clifton \cite{LC11}. They consider what $\epsilon$ means for a hypothetical
adversary, who is trying to discover whether an individual has participated in a
database or not. The core idea is to model the adversary as a Bayesian agent,
maintaining a belief about whether the individual is in the database or not.
After observing the output of a private mechanism, he updates his belief
depending on whether the outcome was more or less likely if the individual had
participated.

As Lee and Clifton show, $\epsilon$ controls how much an adversary's belief can
change, so it is possible to derive a bound on  $\epsilon$ in order for the
adversary's belief to remain below a given threshold. We share the goal of Lee
and Clifton of deriving a bound for $\epsilon$, and we improve on their work.
First, the ``bad event'' they consider is the adversary discovering an
individual's participation in a study. However, by itself, this knowledge might
be relatively harmless---indeed, a goal of differential privacy is to
consider harm beyond reidentification.

Second, and more seriously, the adversary's Bayesian updates (as functions of
the private output of the mechanism) are themselves differentially private: the
distribution over his posterior beliefs conditioned on the output of the
mechanism is nearly unchanged regardless of a particular agent's participation.
In other words, an individual's participation (or not) will usually lead to the
same update by the adversary.  Therefore, it is not clear why an individual
should be concerned about the adversary's potential belief updates when thinking
about participating in the study.

Related to our paper, there are several papers investigating (and each proposing
different models for) how rational agents should evaluate their costs for
differential privacy \cite{Xia13,GR11,NOS12,CCKMV11,LR12}. We adopt the
simplest and most conservative of these approaches, advocated by Nissim, et al.
\cite{NOS12}, and assume that agents costs are upper bounded by a linear
function of $\epsilon$.

Alternatively, privacy (quantified by $\epsilon$) can be thought of as a
fungible commodity, with the price discovered by a market. Li, et al.
\cite{LLMS12} consider how to set arbitrage-free prices for queries.  Another
line of papers \cite{GR11,FL12,LR12,AH12,DFI12,RS12} consider how to discover
the value of $\epsilon$ via an auction, when $\epsilon$ is set to be the largest
value that the data analyst can afford.

\iffull
\section{Conclusion}\label{sec:conclusion}

We have proposed a simple economic model that enables users of differential
privacy to choose the key parameters $\epsilon$ and $\delta$ in a principled
way, based on quantities that can be estimated in practice. To the best of our
knowledge, this is the first comprehensive model of its kind. We have applied
our model in two case studies, and we have used it to explore the surprising
observation that a private study can be cheaper than a non-private study with
the same accuracy. We have discussed ways in which our model could be refined,
but even in its current form the model provides useful guidance for practical
applications of differential privacy.
\fi

\bibliographystyle{abbrv}
\bibliography{epsilon}

\iffull
\appendix
\fi

\section{Protected and unprotected events} \label{app:protected}
In this section, we take a closer look at how to define the space of events in
our model (Section~\ref{sec:interp}).

As alluded to before, not all events are protected by differential privacy---the
probability of some events may become a lot more likely if an individual
participates. In particular, events that can observe an individual's
participation are not protected under differential privacy---this is why we have
defined $\mathcal{E}$ to exclude these events. For a trivial example, the event
``John Doe contributes data to the study'' is very likely if John Doe
participates, and very unlikely if John Doe does not participate.

However, not all cases are so obvious. Consider the following scenario: an
adversary believes that a differentially private study is conducted on either a
population of cancer patients, or a control group of healthy patients. The
adversary does not know which is the case, but the adversary knows that John Doe
is part of the study.

Now, suppose the study provider releases a noisy, private count of the number of
cancer patients in the study. For this answer to be remotely useful, it must
distinguish between the case where all the participants have cancer and the case
where none of the participates do. Hence, the adversary will be reasonably
certain about whether the participants in the database have cancer, and about
whether John Doe has cancer. This seems to violate differential privacy: by
participating in the study, John Doe has revealed private information about
himself that would otherwise be secret. Where did we go wrong?

The key subtlety is whether the adversary can observe John Doe's participation.
Suppose that the released (noisy) count of cancer patients is
$n$, and suppose the bad event John Doe is worried about is ``the adversary
thinks that John Doe has cancer.'' In order for this event to be protected by
differential privacy, it must happen with the same probability whether John Doe
participates or not, whenever the noisy count is $n$.

If the adversary can truly observe John Doe's participation, i.e., he can tell if John
Doe actually participates or not, then clearly this is not the case---if John
Doe participates, the adversary will believe that he probably has cancer, and
if John Doe does not participate, the adversary will not believe this.

On the other hand, if the adversary  merely {\em believes} (but could be
mistaken) that John Doe participated, the bad event {\em is} protected by
differential privacy---if John Doe participates and the adversary discovers that
he has cancer, the adversary would still think he has cancer even if he had not
participated.

This example also illustrates a fine point about the notion of privacy implicit in
differential privacy: while an informal notion of privacy concerns an adversary
correctly learning an individual's secret data, differential privacy deals
instead with the end result of privacy breaches. If John Doe does not participate
but the adversary thinks he has cancer, John Doe should not be happy just because
the adversary did not learn his private data---his insurance premiums
might still increase. In this sense, differential privacy guarantees that he
is harmed nearly the same, whether he elects to participate or not.


\section{The individual's participation decision} \label{app:participate}


In many situations, the data has already been collected, and the individuals may
not have a meaningful choice to participate or not in the study---they may have
already moved on, and it might be difficult to track them down to compensate
them. However, our model assumes the individual has a real choice about whether
to participate or not.

To get around this problem, we could imagine running a thought experiment to
calculate how much we would have had to pay each student to incentivize them to
participate if they really had a choice. Since in this thought experiment
participants have a choice, our model applies. If the required compensation is
small, then the expected harm to any student is not very high, and we can run
the study. 

However, there is yet another problem.  Recall that our model assumes the study
will be run, and compensates individuals for their marginal increase in expected
harm when participating. In general, the increase in cost for participating in a
study compared to not running the study at all can be high.\footnote{%
  This corresponds to applying the private mechanism to two databases: the real
  one, and an empty one (representing the case where the study is not run).
  These databases are far apart, so the potential increase in harm could be
  large.
}
Since our study will not be run if the harm to the individuals is high, should
the individuals demand more compensation (i.e., is their privacy harm higher
than predicted by our model)?

The answer turns out to be no. We first estimate the costs, and then decide
whether or not to run the study. If we do, we ask individuals if they want to
participate in exchange for compensation (in our thought experiment). The point
is that when the individuals are given the (hypothetical) choice, we have
already decided that the study will happen. Thus, they should be compensated for
their marginal harm in participating, and no more.

\end{document}